\documentstyle[12pt,epsfig]{article}
\def\etal{{\it et al.}}

\def\ie{{\it i.e. }}

\def \lsim{\lower.8ex\hbox{$\sim$}\kern-.8em\raise.45ex\hbox{$<$}\;}
\def \gsim{\lower.8ex\hbox{$\sim$}\kern-.8em\raise.45ex\hbox{$>$}\;}
\begin{document}
\begin{titlepage}
\bigskip
\bigskip
\begin{center}
{\Large \bf New possibilities of old soft  pomeron in DIS}
\vskip 0.5cm
{\bf P. Desgrolard}${}^{\dag}$,\\
{\it Institut de Physique Nucl\'eaire de Lyon, IN2P3-CNRS et Universit\'{e}
Claude Bernard,\\
43 boulevard du 11 novembre 1918, F-69622 Villeurbanne Cedex, France\\}
{\bf A. Lengyel}${}^{\dag\dag}$,\\
{\it Institute of Electronic Physics, National Academy of Sciences of
Ukraine,\\
294015 Uzhgorod-015, Universitetska 21, Ukraine\\}
{\bf E. Martynov}${}^{\ddag}$,\\
{\it N.N. Bogoliubov Institute for Theoretical Physics, National Academy of
Sciences of Ukraine,\\
252143, Kiev-143, Metrologicheskaja 14b, Ukraine\\}
\footnotetext{${}^{\dag}$ \hskip 5pt E-mail: desgrolard@ipnl.in2p3.fr}
\footnotetext{${}^{\dag\dag}$ E-mail: sasha@len.uzhgorod.ua}
\footnotetext{${}^{\ddag}$ \hskip 5pt E-mail: martynov@gluk.apc.org}
\vskip 1.5cm
{\bf Abstract}
\end{center}

A traditional Regge model with a
$Q^2$-independent Pomeron intercept closed (or equal) to
one is constructed in order to describe the available data on the proton
structure function. A Dipole Pomeron model which does not explicitly violate
unitarity is developed and investigated. An excellent agreement with
the 1209 data is found ($\chi^2/{dof}=1.11$) in the whole kinematical
domain
investigated by experiments.
A comparison of the model with already existing ones is
made. The $x-$, $Q^2-$slopes and the effective
intercept are discussed as $Q^2$ and $x$ functions.
\end{titlepage}

\vskip 1.cm

{\Large \bf Introduction}
\vskip 0.3cm

The smooth transition between non-per\-tur\-ba\-ti\-ve (soft
 pomeron and Reg\-ge\-ons) and perturbative (QCD evolution, hard  pomeron)
behaviour was studied and discussed in a number of original papers
and reviews ({\it e.g.} \cite{BGJPP,Lev}). Important questions
remain, however,
unresolved about the kinematical region (in $Q^2,x$-plane) in which
the above approaches can be applied as well as about the region of their
interference. In particular:
\begin{itemize}
\item[i)] How large are
the corrections to the DGLAP \cite{dglap} evolution equation where the
$Q^2$-evolution is usually considered in the leading $log Q^2$
approximation \cite{L1,L2}~?
\item[ii)] How  large are the corrections
to BFKL- or hard- pomeron \cite{bfkl}~? What is their influence on
the structure of the singularities in the $j$-plane, on the
position of the rightmost singularity and on its intercept
\cite{L1,L2,FL}~?
\item[iii)] What do we know about unitarity in lepton-hadron Deep
Inelastic Scattering (DIS)? How important are the shadowing corrections
(SC) to the hard  pomeron at small values of $x$ \cite{L1,L2,FDL,GLM}~?
Does the Froissart-Martin bound for hadronic total cross-sections
remain valid for $\gamma p$-interaction?  \item[iv)] What are the
domains in $Q^2$ and $x$ where a Regge description of the structure
functions (SF) can be applied~?
\end{itemize}

Detailed discussion may be found in the above quoted papers and
references therein; here we briefly review the main conclusions
known so far.

\begin{itemize}
 \item[I)]
The experimental data on the deep inelastic SF are successfully
described by the DGLAP evolution equation without any new
ingredients \cite{GRV,MRS,CTEQ} providing an initial structure
function $F_2^{(0)}\propto x^{-\omega_0}$ at $x\to 0$ with
$\omega_0\approx 0.2
- 0.3$. We note that an important point in this approach is the choice
of starting $Q^2$ value at which the input is defined. This value
(usually $\sim 1 -4$ GeV$^2$) is taken on a phenomenological ground and
is justified {\it a posteriori}. At the same time in the
HERA kinematical region the next order corrections are believed to be
important \cite{L1,L2}. From this point of view a good agreement of
perturbative results with the data can be considered more strange
than natural.
\item[II)]
As shown recently \cite{FL}, the
correction $\delta \omega$ to the "Born" intercept of the BFKL Pomeron
$$\alpha_{\cal P}^{(0)}(0)-1=\omega_0=3N_c(\alpha_s/\pi )\ln2 \approx
0.397,$$ calculated in the first order in $\alpha_s(\approx 0.15)$, is
large and negative. More precisely, in accordance with estimates made in
\cite{FL}
$$\omega=\omega_0+\delta \omega \approx 0.0747 \quad \mbox{if} \quad
\alpha_s=0.15\ ,$$ $$\omega \approx0.214 \quad \mbox{if} \quad
\alpha_s=0.081\ . $$
The authors of \cite{FL} conclude that the BFKL Pomeron
and
its next to leading approximation can be used only for rough
estimates rather than for "precise" phenomenology.
\item[III)]
Quantitative estimates of unitarity, shadowing corrections to
structure functions as well as to parton distribution functions
depend on additional assumptions within a specified procedure of
unitarization. However, numerical estimates of SC originated from
short distances show \cite{FDL,HLS} that these effects are not
small though they do not change qualitatively the behaviour of
structure functions at least at moderate $Q^2$ and when $x\lsim
10^{-2}$. As for the unitarity condition in DIS, there is a common
belief that the Froissart-Martin bound can not be proved for a
process including "external" off-mass-shell particles.
Nevertheless, as shown in \cite{Petr}, some restrictions on the
values of the intercepts can be obtained on the ground of
unitarity. We shall return to this subject below.
\item[IV)]
In accordance with a widely accepted point of view, a
soft contribution to the proton structure function $F_2^p$ dominated by the
pomeron works only at small $Q^2$.  The basis of this belief is that at
fixed $Q^2$ (larger than a few units of GeV$^2$) a {\it simple} fit
(without any subasymptotic term which could be important here)
gives $F_2^p\sim x^{-\delta}$ where $\delta \approx 0.2 - 0.4$ and
$\delta $ is rising with $Q^2$.
\end{itemize}

The visible dependence of $\delta $ on $Q^2$ was used in the CKMT
\cite{CKMT} and ALLM models \cite{ALLM,AL} where a  pomeron with an
intercept
depending on $Q^2$ was introduced.

Besides, a two pomerons model \cite{land94} and other models were
proposed which smoothly interpolate between a soft and a hard
$Q^2-$dependence \cite{BGP} or combine these behaviours \cite{AY}.
Of course, such a picture for the pomeron does not correspond to a
true Regge singularity. It contradicts the main properties of
simple Regge poles, namely factorization and universality. Rather
it can be considered as an effective contribution taking into
account possible multipomeron exchanges. It would be interesting
and important to find a justification by directly summing the
multipomeron terms for example by an eikonal or a quasieikonal
method. However it has not been done yet.

Let us come back to the result of \cite{Petr}. Because of
unitarity, two kinds of singularity of the amplitude are possible:
Regge singularities, those trajectories $\alpha (t)$ are,
naturally, $Q^2$-independent; and Renormalization Group
singularities which can depend on $Q^2$. From unitarity, the
inequality
\begin{displaymath}
\alpha (Q^2)-1 < \alpha (0)-1
\end{displaymath}
follows.

In spite of the above mentioned belief on the validity of a
"non-hard pomeron" description of DIS data, restricted to small and
moderate $Q^2$, many models of a soft pomeron contribution to
$F_2^p$ \cite{AY,M,JMP,BH,SS,DGJLP} were constructed and proved to
be successful at small $x$ ($\lsim 10^{-2})$ and in a wide region
of $Q^2$.

It is interesting to note that in most of them
\begin{displaymath}
F_2^p(x,Q^2)_{\longrightarrow \atop x\to 0} f(Q^2)\ln (\frac{1}{x})
\cong f(Q^2)\ln W^2,\qquad  \mbox{where}\qquad
W^2=Q^2(\frac{1}{x}-1)+m_p^2
\end{displaymath}
($m_p$ is the proton mass). Such a behaviour corresponds exactly to the
contribution of a double $j$-pole $f(Q^2)/(j-1)^2$ to the partial amplitude
of $\gamma^*p\longrightarrow \gamma^*p$, where $f(Q^2)$ is the residue function
of the given reggeon (or pomeron). In hadronic models for elastic
scattering, this singularity is known as a Dipole Pomeron (DP) with a
trajectory $\alpha_{\cal P}(t)$, having unit intercept,
$\alpha _{\cal P}(0)=1$.
As was shown in \cite{DGLM,DGM} the Dipole Pomeron model gives rise to the
"best" description (in sense of $\chi^2$) of the experimental data on the
total cross section and $\rho-$ratio for nucleon-nucleon scattering
($pp$ and $\bar pp$) as well as for meson-nucleon.

All this leads support to our present efforts to answer the following question.
Is it possible, keeping a pure Regge picture, to extend the area of
validity of the soft  pomeron~?

Taking into account the results of different "soft" models
successfully applied at small \cite {M,JMP,SS,DL} and at moderate
$Q^2$ \cite{M,JMP,BH,SS} one expect that the main difficulty should
be the description of the data at large $x$ rather than at large
$Q^2$.

\vskip 1.cm

{\Large \bf
1 The model}
\vskip 0.3cm

As natural in a Regge approach, we deal with amplitudes and cross-sections
rather than with structure functions. Therefore,  we start from the
expression connecting the transverse cross-section
$\sigma_T(W,Q^2)$ for the ($\gamma^*,p$) process to the proton SF
$F_2^p(x,Q^2)$
\begin{equation}\label{1} \sigma _{T}^{\gamma ^{*}p}(W,Q^2)=\frac{4\pi
^{2}\alpha
}{Q^2}\frac{1}{1-x}(1+\frac{4m_{p}^{2}x^{2}}{Q^{2}})\frac{1}{1+R(x,Q^2)}F_2(
x,Q^2)\ ,
\end{equation}
where we recall the negative squared four-momentum transfer carried
by the virtual photon $Q^2$, the Bj\"orken variable $x$ and the
center of mass energy of the $\gamma^*p$ system $W$ obey the
condition
\begin{displaymath}
W^2=Q^2\frac{1-x}{x}+m_p^2\ ;
\end{displaymath}
here $\alpha$ is the fine structure constant and
\begin{displaymath}
R(x,Q^2)=\frac{\sigma _{L}(x,Q^2)}{\sigma _{T}(x,Q^2)}.
\end{displaymath}

Unfortunately, the longitudinal cross-section $\sigma_L$ is poorly known
and one only knows that $R(x,Q^2)$ is small
(at least at small $Q^2$ and $x$). In what follows we approximate
$$R(x,Q^2)=0,$$
(\ie the total and transverse cross-sections are supposed to be the same).
Thus, we use the expression
\begin{equation}\label{2}
F_2^p(x,Q^2)=\frac{1}{4\pi ^2\alpha}a(x,Q^2)\sigma _{T}^
{\gamma ^{* }p}(W,Q^2) \ ,
\end{equation}
with
\begin{equation}\label{3}
a(x,Q^2)=\frac{Q^2(1-x)}{1+4m_p^2x^2/Q^2}.
\end{equation}

From the optical theorem
\begin{equation}\label{4}
\sigma _{T}^{\gamma ^{*}p}(W,Q^2)=\Im m \left[A(W^2,t=0,Q^2)\right]\ ,
\end{equation}
we normalize the elastic scattering amplitude $A(W^2,t,Q^2)$ with the
external particles $\gamma^*$ off mass shell. Thus
\begin{equation}\label{5}
F^p_2(x,Q^2)=\frac{1}{4\pi ^{2}\alpha }a(x,Q^2)\ \Im m A(W^2,t=0,Q^2).
\end{equation}
The Regge model says nothing about the $Q^2$-dependence of the amplitude.
The data (Fig.~1), however, suggest a power-like
decrease of the cross-section at fixed $W$.
\vskip 0.5cm
\begin{center}
\epsfig{figure=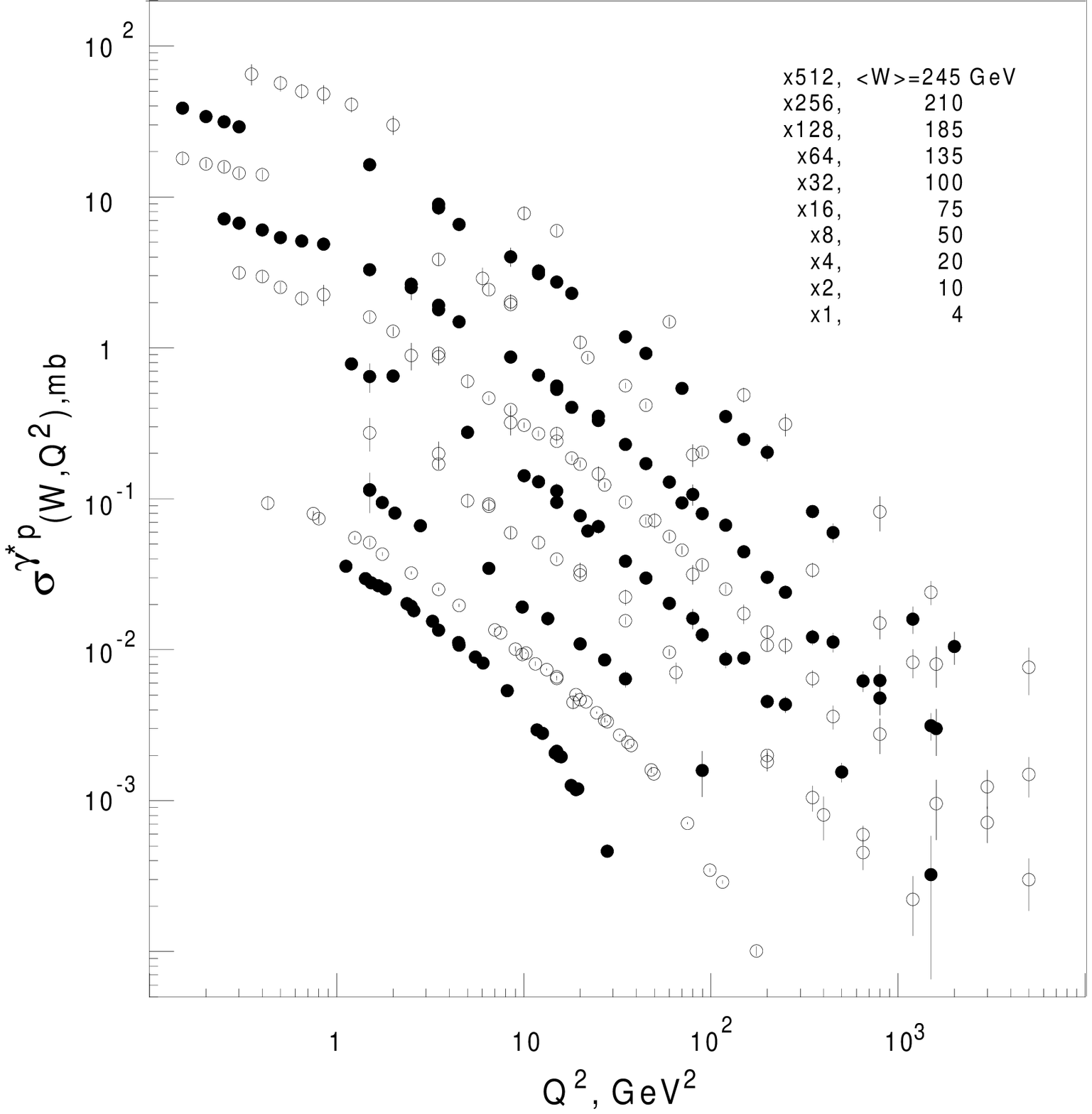,width=12cm}
\end{center}
\vskip 0.5cm
{\bf Fig.~1.}
 Experimental data for $\sigma_T^{\gamma ^*p}$ versus $Q^2$ at
fixed $W$'s showing a power decrease at high $Q^2$.
\vskip 0.5cm
In the amplitude we take into account the contributions of a
pomeron and of a secondary reggeon ($f$-reggeon).
\begin{equation}\label{6}
A(W^2,t=0,Q^2)=P(W^2,Q^2)+F(W^2,Q^2).
\end{equation}
Strictly speaking, an $a_2$-reggeon should also contribute to the
$\gamma^*p$-amplitude (and is expected to be important at low
energy). We do not include it in the fit to avoid extra free
parameters and therefore we fit the model only in the region $W\ge
3$ GeV.

The f-reggeon contribution is written
\begin{equation}\label{7}
F(W^2,Q^2)=iG_f(Q^2)\bigg ( -i\frac{W^{2}}{m_{p}^{2}}\bigg )^
{\alpha _{f}(0)-1}\left(1-x \right)^{B_f(Q^2)},
\end{equation}
where we take
\begin{equation}\label{8}
G_f(Q^2)=\frac{g_f}{\left(1+Q^2/Q_{f}^2 \right)^{D_f(Q^2)}},
\qquad D_f(Q^2)=d_{f\infty}+\frac{d_{f0}-d_{f\infty}}{1+Q^2/ Q_{fd}^2},
\end{equation}
\begin{equation}\label{9}
B_f(Q^2)=b_{f\infty}+\frac{b_{f0}-b_{f\infty}}{1+Q^2/Q^2_{fb}}.
\end{equation}
As for the pomeron contribution, we take it in the form
\begin{equation}\label{10}
P(W^2,Q^2)=P_1+P_2,
\end{equation}
with
\begin{equation}\label{11}
P_1=iG_1(Q^2){\cal P}(W)(1-x)^{B_1(Q^2)},
\qquad P_2=iG_2(Q^2)(1-x)^{B_2(Q^2)},
\end{equation}
\begin{equation}\label{12}
G_i(Q^2)=\frac{g_i}{\left(1+Q^2/Q_{i}^2 \right)^{D_i(Q^2)}},
\qquad D_i(Q^2)=d_{i\infty}+\frac{d_{i0}-d_{i\infty}}{1+Q^2/ Q_{id}^2},
\qquad i=1,2,
\end{equation}
\begin{equation}\label{13}
B_i(Q^2)=b_{i\infty}+\frac{b_{i0}-b_{i\infty}}{1+Q^2/Q^2_{ib}}, \qquad
i=1,2.
\end{equation}

We would like to comment the above expressions. In spite of their
(apparently) cumbersome form, they follow from a direct
generalization of the simplest parameterization of factors
$$G(Q^2)=\frac{g}{(1+Q^2/Q_0^2)^d}\qquad \mbox{and}\qquad (1-x)^b$$
with constant $d$ and $b$ in each term of the
$\gamma^*p$-amplitude. A fit to experimental data shows that the
values of d and b should depend on $Q^2$.

Various models of the pomeron may be considered (via ${\cal P}(W)$), {\it
e.g.}
\begin{itemize}
\item {\bf Dipole Pomeron (DP)}
\begin{equation}\label{14}
{\cal P}(W)=\ln (-i\frac{W^2}{m_p^2}),
\end{equation}
 \item {\bf Supercritical  Pomeron (SCP)}
\begin{equation}\label{15}
{\cal P}(W)=\bigg (-i\frac{W^2}{m_p^2}\bigg )^{\alpha _{P}(0)-1},
\end{equation}
\item {\bf "Generalized" Pomeron}
\begin{equation}\label{16}
{\cal P}(W)=\ln^{\mu } (-i\frac{W^2}{m_p^2}),\qquad 0\le \mu \le 2.
\end{equation}
\end{itemize}

We only investigate the first two models with a soft pomeron (here
with an intercept close to one). In the DP model, the intercept of
the pomeron is $\alpha_{\cal P}(0)=1$, while in the SCP model, the
pomeron intercept is fixed at its "world value" $\alpha_{\cal
P}(0)=1.0808$. Note that our SCP model is a generalization of the
model by Donnachie and Landshoff \cite{DL}~: we add in the
amplitude a single term with a unit intercept.

The parameters must obey some restrictions to avoid unphysical situations
(for example,
the cross-section might become negative if we do not constrain
$d_{2\infty}\geq d_{1\infty}$).
These restrictions were taken
into account when fitting (6-13) to the available experimental data.

In Table 1 we show details on the set of experimental data
used for the determination
of the parameters to analyze the properties of the model.
\vskip 1.5cm
\noindent
{\bf Table~1.} Experimental data used in the fit of DP and SCP
models. Distribution of the partial $\chi^2$'s for each subset of
data is illustrated for the DP model.
\vskip 0.5 cm
{\small
\begin{displaymath}
\begin{array}{|c|c|c|c|}
\hline
{\rm Experiment,}   &  {\rm Number}   & {\rm  Reference}
& \chi^2 \,\,{\rm in}\\
{\rm observable\, quantity} &{\rm of\, points} & & {\rm DP \,\,model}\\
\hline
{\bf {\sigma}}_T \,\,(W\geq 3\ {\rm GeV}) & 99 & Durham Data Base,
& 122.79 \\
 & & Zeit.\ Phys.{\bf C63(1994)391} &\\
  & & Zeit.\ Phys.{\bf C69(1995)27} &\\
\hline
{\rm \bf F_2^p} & & &\\
H1 & 193 & Nucl.\ Phys. {\bf B470}(1996)3
&108.05 \\
 H1 & 44 & Nucl.\ Phys. {\bf B497}(1997) 3 & 39.18\\
 H1 & 93 & Nucl.\ Phys. {\bf B439}(1995)471 & 67.42 \\
 ZEUS & 188 & Zeit.\ Phys.{\bf C72}(1996)399 & 233.70\\
 ZEUS & 34 & Phys.\ Lett. {\bf B407}(1997)432 & 22.20\\
 BCDMS, (W\geq 3\ {\rm GeV}) & 175 &
Phys.\ Lett. {\bf B223}(1989)485 & 285.40 \\ NMC, (W\geq 3\ {\rm
GeV}) & 156 & Nucl.\ Phys. {\bf B483}(1997)3 & 175.40\\
 E665 & 91 & Phys.\ Rev. {\bf D54}(1996)54 & 95.40 \\
 SLAC, (W\geq 3\ {\rm GeV})
& 136 & Phys. Lett. {\bf B282}(1992) 475& 167.50 \\

& & SLAC-PUB\, 357(1990) & \\\hline {\rm Total} & N_p=1209 & &
\chi^2/d.o.f.=1.11\,\, \\
\hline
\end{array}
\end{displaymath}
}
\vskip 1. cm

{\Large \bf 2 Results and discussion}

\vspace{0.5cm}

{\bf 2.1 Fit of the data}

We performed a fit of the experimental data with the two models:
the Dipole Pomeron and the Supercritical Pomeron. In the DP model,
the Reggeon intercept was fixed at the value $\alpha_f(0)=0.804$
obtained from hadronic reactions \cite{DGLM,DGM}. As regards the
SCP model only the intercept of Pomeron was fixed at the value
$\alpha_{\cal P}(0)=1.0808$ though an another value of
$\alpha_{\cal P}(0)$ was obtained in \cite{DGLM,DGM}.

The corresponding $\chi^2$ and the fitted parameters are given in
Tables~1-2.

Both models describe well the data. In practice, they give plots
which coincide in the region of the fitted experimental data; they
become different only in the very far asymptotics or, as
anticipated, at low $W$-values. We plot $\sigma_T^{\gamma
^{*}p}(W,Q^2)$ in Figs.~2a-b and $F_2^p(x,Q^2)$ in Figs.~3a-d; our
DP results are compared to ALLM ones \cite{AL}, recalculated after
corrections of a few misprints \cite{HA}. In what follows we
concentrate mainly on a discussion of the DP model.

\begin{table}
\noindent
{\bf Table~2.} Parameters obtained in the Dipole Pomeron
model and in the Supercritical Pomeron model.
\vskip 0.5cm
{\small
\begin {center}
\begin{tabular}{|c|l|l|}
\hline
 & \hskip 0.6cm DP model & \hskip 0.6cm SCP model   \\
\hline
 Parameters    & \hskip 0.9cm Value   & \hskip 0.9cm Value  \\ \hline
 $P_1$-term  & & \\ \hline
$\mu$ & .10000E+01 (fixed)& \\ $\alpha_{\cal P}(0)$
&.10000E+01(fixed) &
.10808E+01 (fixed) \\ $g_1(mb)$ & .21898E-01 & .10295E+00 \\
$Q_1^2(GeV^2$ &
.15400E+02 & .88709E+01 \\ $Q_{1d}^2(GeV^2)$ & .17852E+01 &
.15329E+01 \\ $Q_{1b}^2(GeV^2)$ & .33435E+01 & .99243E+01 \\
$d_{1\infty}$ & .13301E+01 & .13026E+01 \\ $d_{10}$ & .14370E+02 &
.89733E+01 \\ $b_{1\infty}$ & .21804E+01 & .27830E+01 \\ $b_{10}$ &
.42596E+01 & .42832E+01 \\
\hline
$P_2$-term & & \\
\hline $g_2(mb)$ & -.99050E-01 & -.78055E-01 \\
$Q_2^2(GeV^2$ & .34002E+02 & .20269E+02 \\ $Q_{2d}^2(GeV^2)$ &
.12327E+01 & .22877E+01\\
$Q_{2b}^2(GeV^2)$ & .20702E-01 &
.21626E-01 \\
$d_{2\infty}-d_{1\infty}$ & .00000E+00 (fixed) & .00000E+00
(fixed) \\
$d_{20}$ & .22607E+02 & .72161E+01 \\
$b_{2\infty}$ &
.24686E+01 & .30767E+01 \\ $b_{20}$ & .17023E+03 & .25000E+03 \\
\hline
 $F$-term  & & \\
 \hline
$\alpha_f(0)$ & .80400E+00 (fixed)& .71369E+00 \\
$g_f(mb)$ &.29065E+00 & .18189E+00 \\
$Q_f^2(GeV^2$ & .29044E+02 & .22469E+02\\
$Q_{fd}^2(GeV^2)$ & .54462E+00 & .13003E+00 \\
$Q_{fb}^2(GeV^2)$ & .20656E+01 & .49263E+01 \\
$d_{f\infty}$ &.13554E+01 & .12940E+01 \\
$d_{f0}$ & .75127E+02 & .22533E+03 \\
$b_{f\infty}$ & .27239E+01 & .32140E+01 \\
$b_{f0}$ & .64713E+00 & .00000E+00(fixed) \\
\hline
$\chi^2/d.o.f.$ & 1.11 & 1.15 \\
\hline
\end{tabular}
\end{center}
}
\end{table}
\vskip 0.5cm

{\bf 2.2 ($\gamma,p$) cross-section (at $Q^2=0$)}.

In the DP model, from (5-14) we obtain
\begin{equation}\label{17}
\sigma_T^{\gamma p}(W)=g_1\ln \bigg ( \frac{W^2}{m_p^2}\bigg )+
g_2+g_f\cos (\frac{\pi }{2}(\alpha _f(0)-1))\bigg ( \frac{W^2}{m_p^2}
\bigg )^{(\alpha _f(0)-1)}.
\end{equation}
The existing data on the cross section for a real photon-proton
interaction are not precise enough to determine unambiguously the
coupling constants $g_1, g_2, g_f$ and the intercept $\alpha_f(0)$
(this is why we fixed the f-reggeon intercept). The behaviour (mild rise) of
$\sigma ^{\gamma p}_T(W)$ is shown in Fig.2a.

\begin{center}
\epsfig{figure=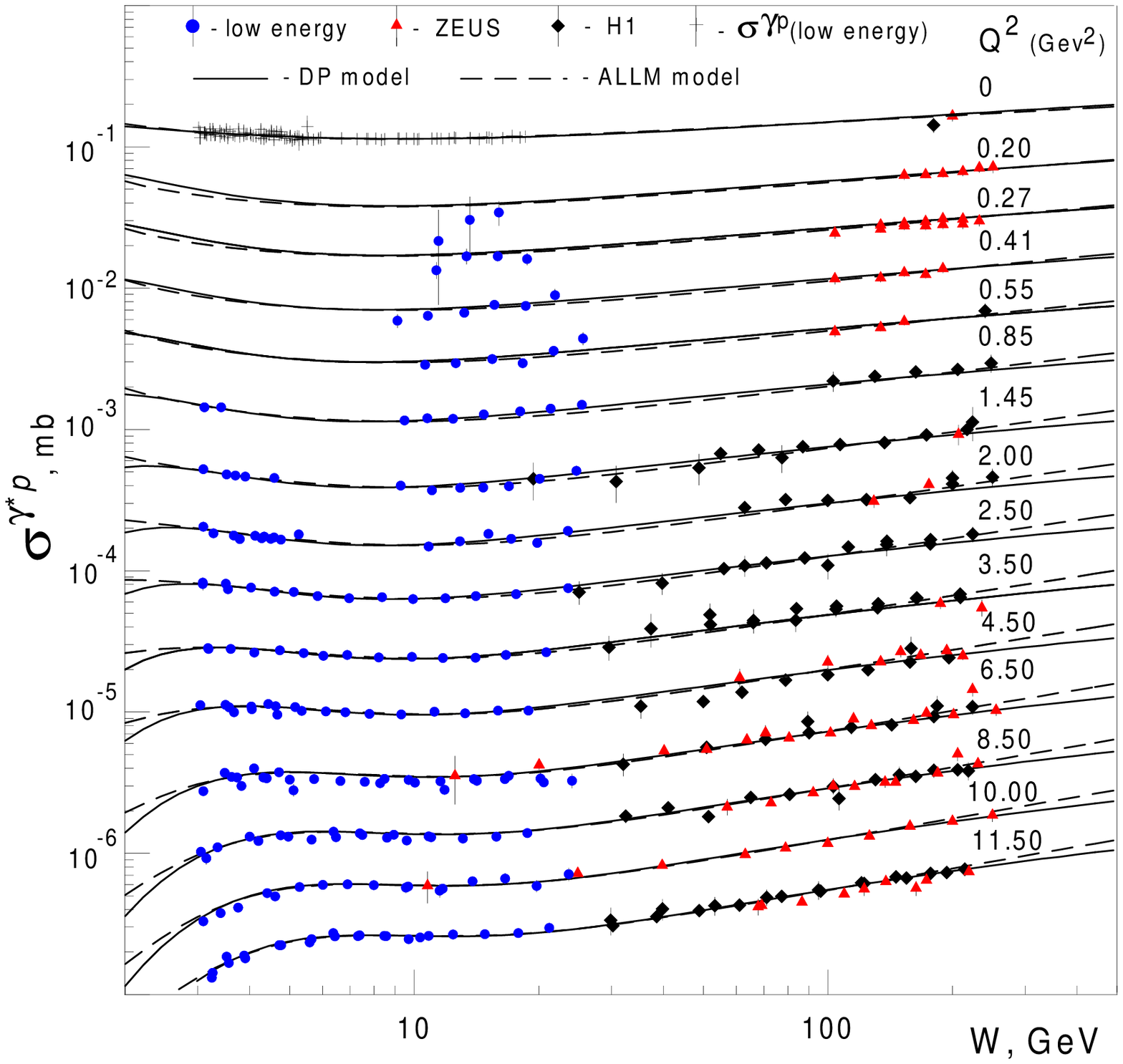,width=16cm}
\end{center}
\vskip 0.5cm
{\bf Fig.~2a.}
 Experimental data for $\sigma ^{\gamma ^*p}(W,Q^2)$
at low $Q^2$ and description in our model and ALLM [17]. A factor 
of 2**(k-1) for each curve is omitted (k is the number of the curve 
starting from the top). Other notations are given in the figure.
\vskip 0.5cm

\begin{center}
\epsfig{figure=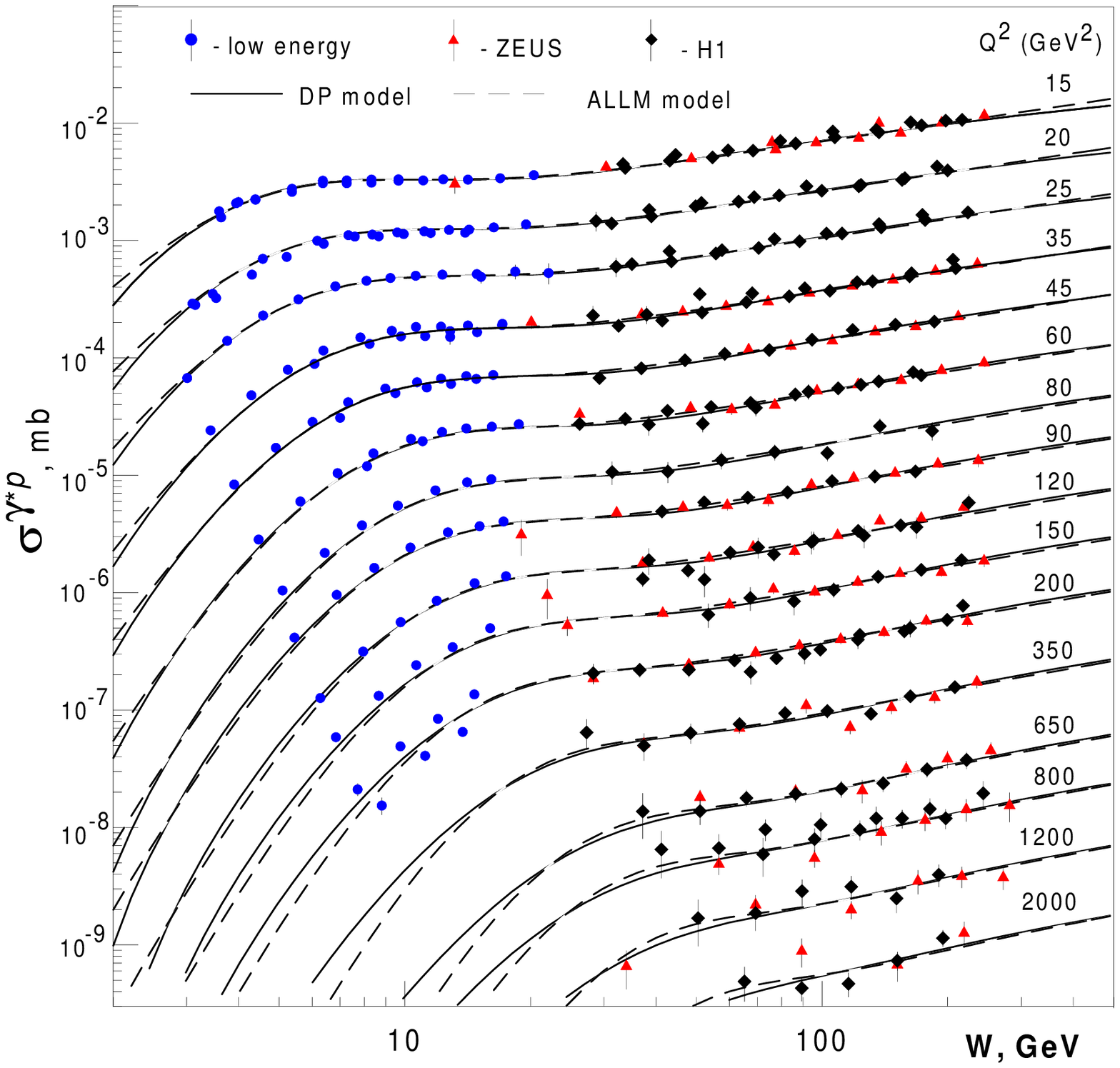,width=16cm}
\end{center}
\vskip .5cm
{\bf Fig.~2b.}
 Same as in Fig.~2a for intermediate and high $Q^2$.
\begin{center}
\epsfig{figure=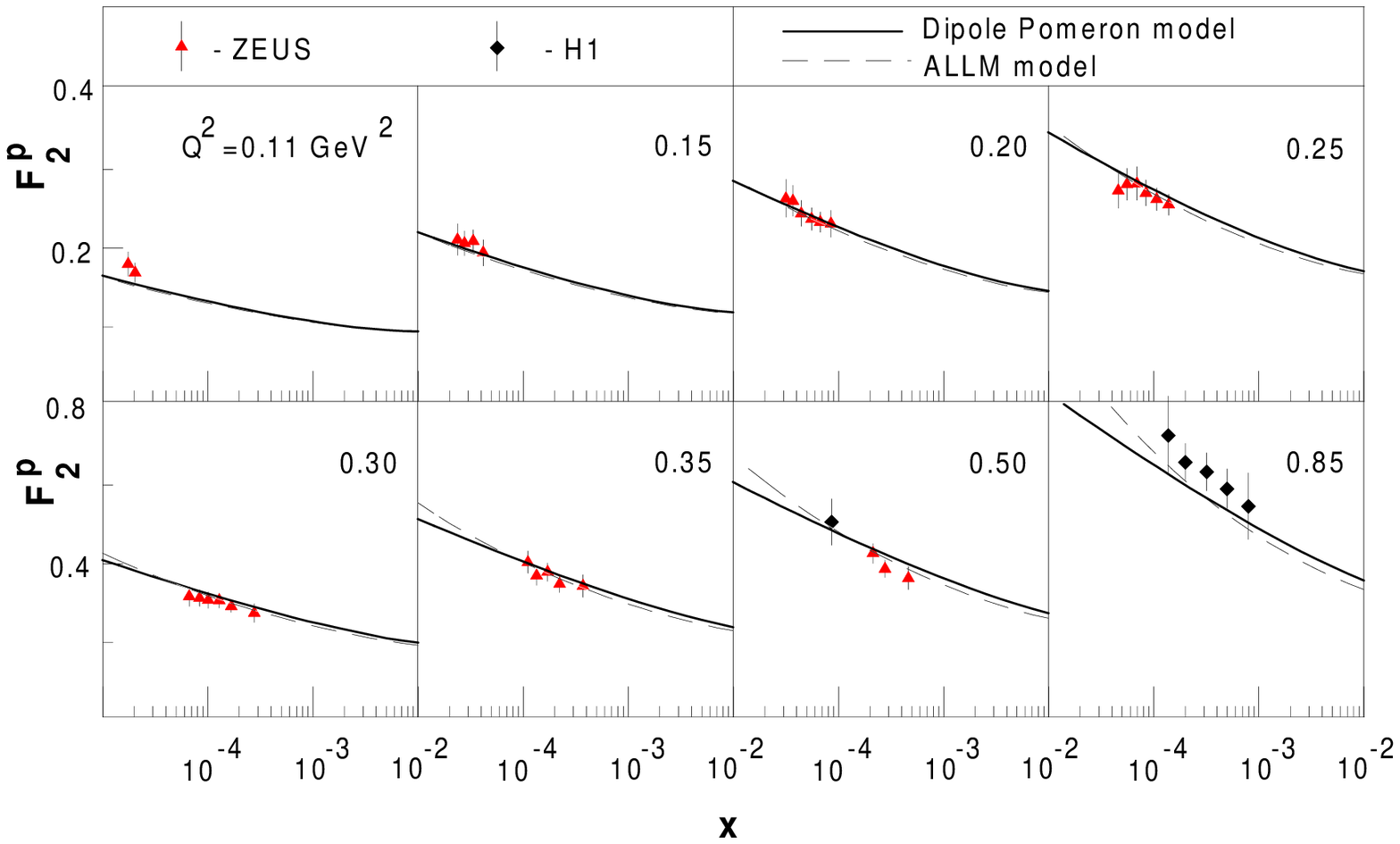,width=15cm}
\end{center}

\begin{center}
\epsfig{figure=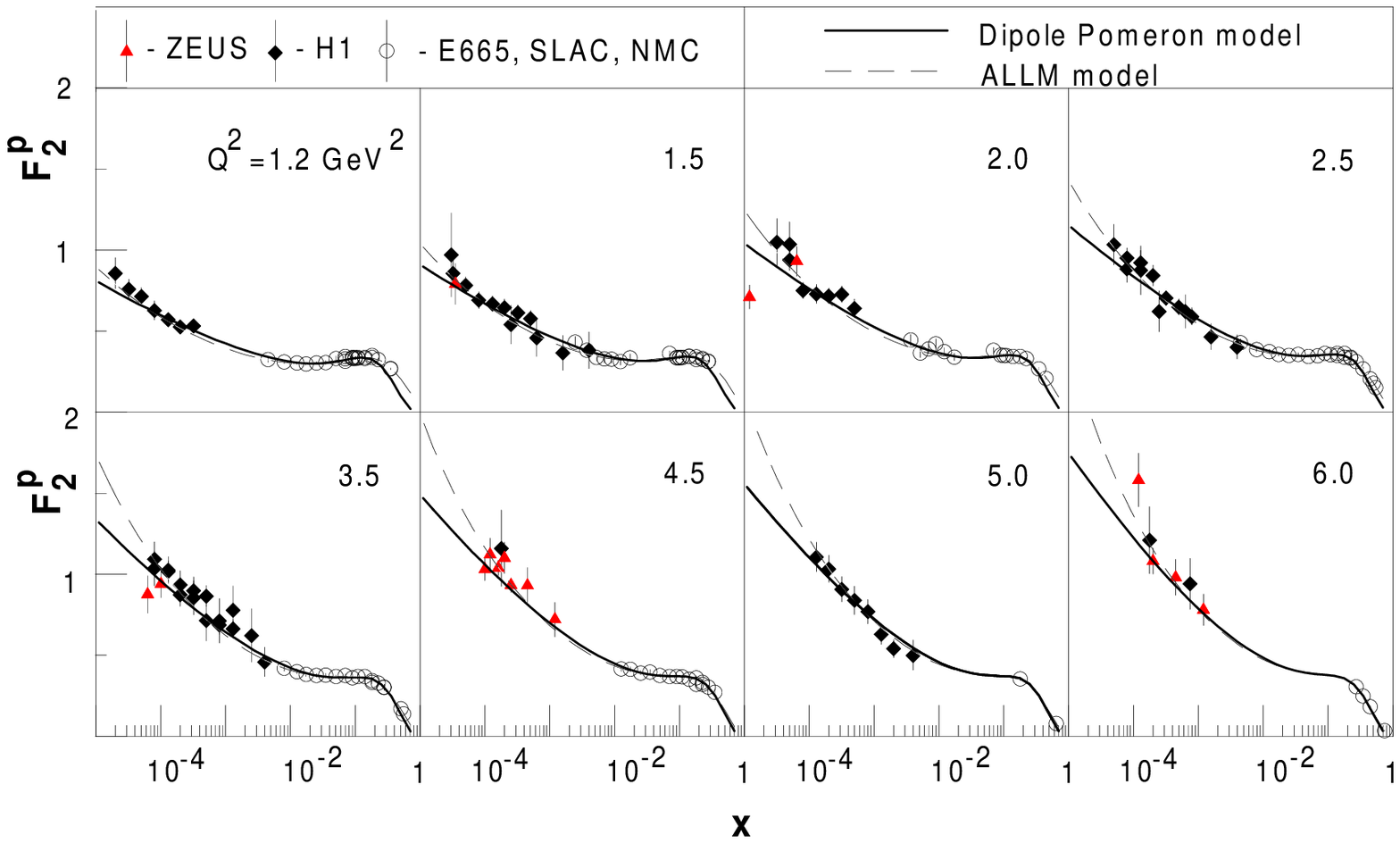,width=15cm}
\end{center}
\vskip .5cm
{\bf Figs.~3a,b.}
 Experimental data for the proton structure function
$F_2^p(x,Q^2)$ at low $Q^2$ and predictions in the Dipole Pomeron
model and in ALLM model. All notations are given in the figure.
\vskip 1.5cm
\begin{center}
\epsfig{figure=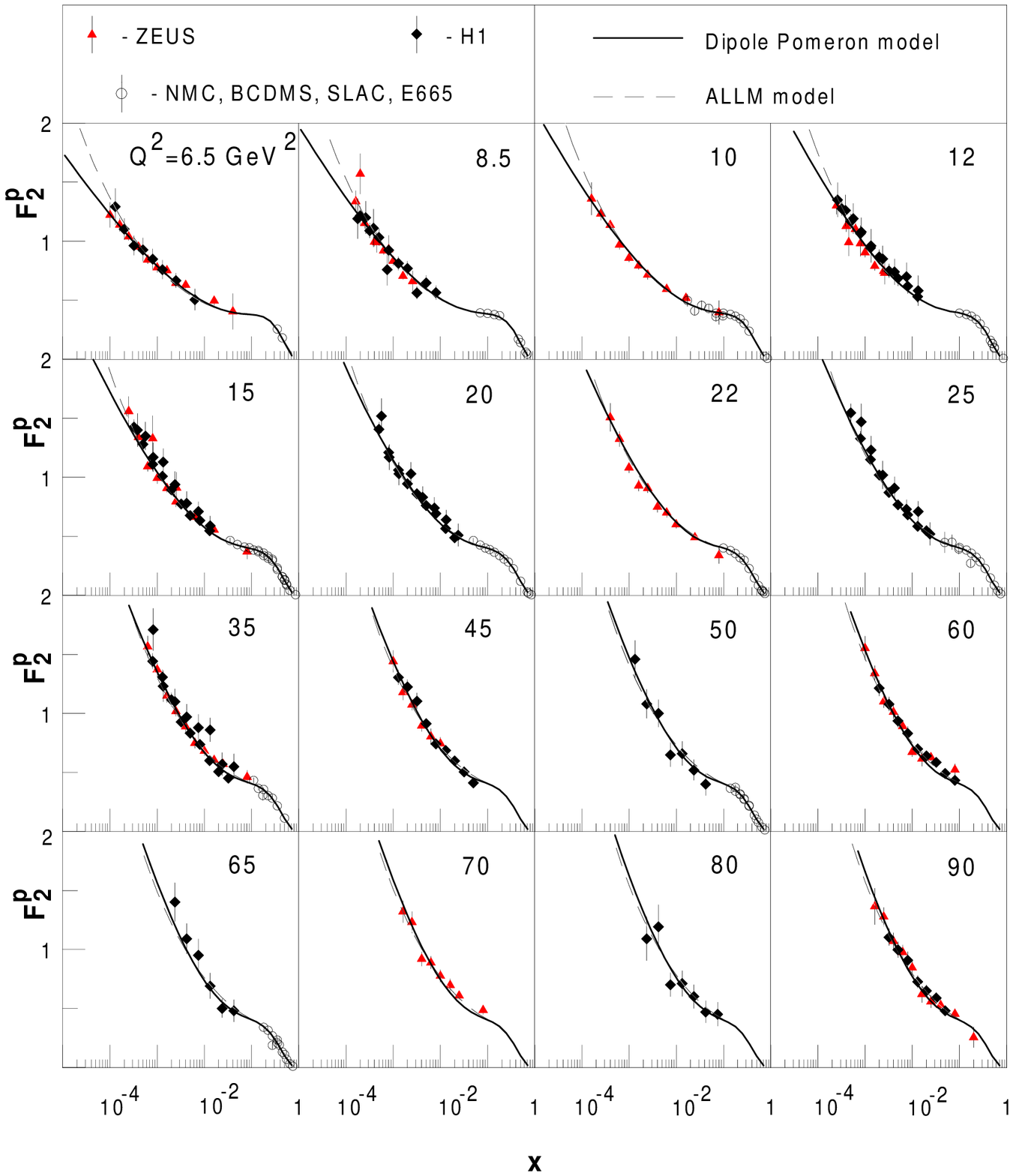,width=16cm}
\end{center}
\vskip .5cm
{\bf Figs.~3c.}
 Same as in Fig.3a,b at intermediate $Q^2$.
\vskip 1.5cm

\begin{center}
\epsfig{figure=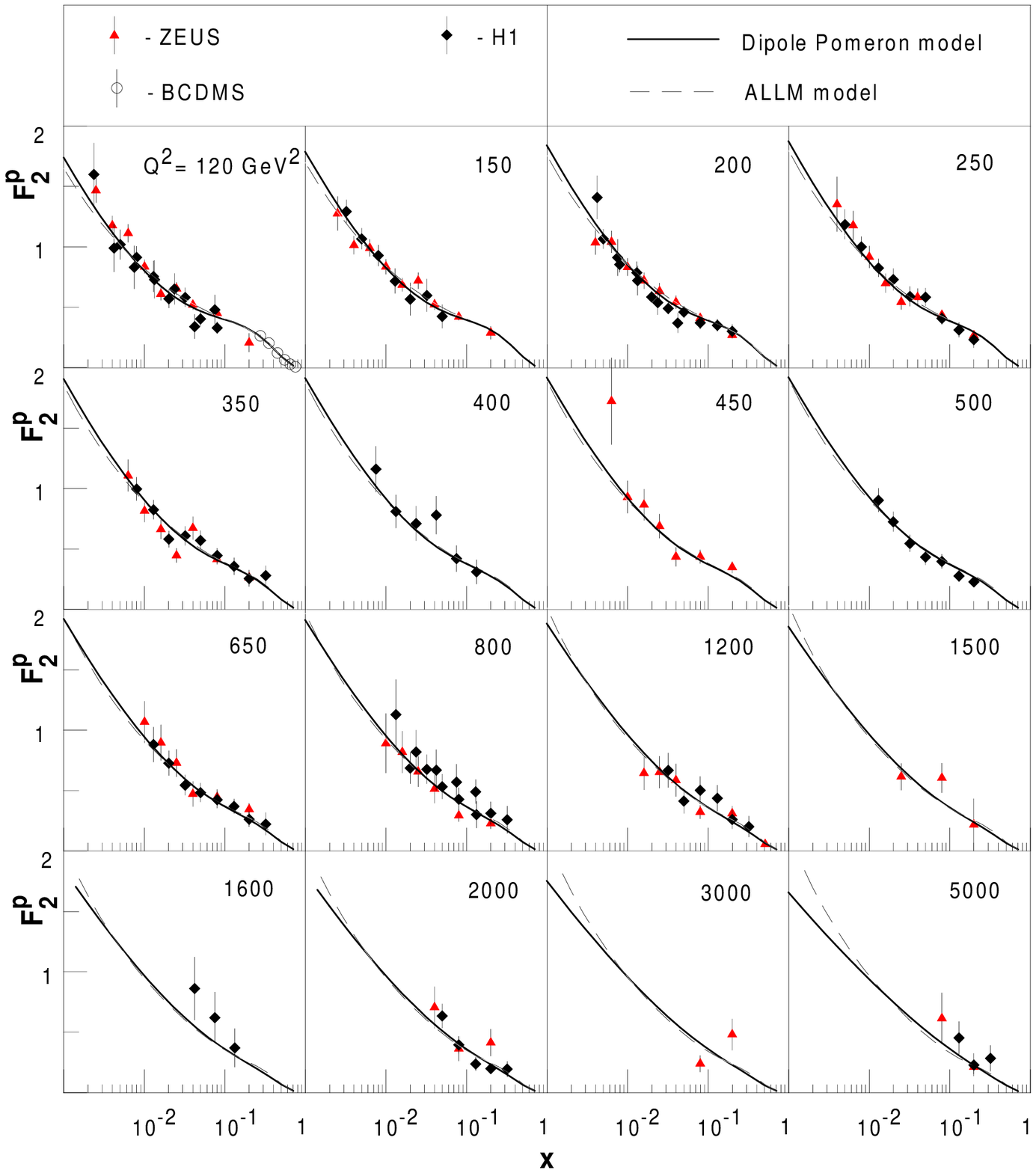,width=16cm}
\end{center}
\vskip .5cm
{\bf Figs.~3d.} Same as in Fig.3c at high $Q^2$.
\vskip 1.5cm

We would like to comment on the properties of the models we
investigated at $Q^2=0$. As usual in a multiparametric problem
there are several minima of $\chi^2$. We found one of them with
$\chi^2/d.o.f.=1.07$ which is significantly better than in the one
given here. In this solution, which we did not retain, the values
of $\sigma_{T}^{\gamma^*p}\sim 0.19$ mb at the HERA energies are
slightly higher than the experimental points. At the same time as
noted in \cite{Bauer} an extrapolation of the ZEUS BPC data to
$Q^2=0$ gives exactly this value of $\sigma_{T}^{\gamma^*p}$.

\bigskip

{\bf 2.3 Partial contributions to the (${\gamma^*, p}$) cross section}.

Let us remark about the negative sign found for the parameter $g_2$ (see
Table 2). The same situation takes place for $pp$ and $\bar pp$
cross-sections \cite{DGLM,DGM}, where a negative contribution plays
an important role for a good description of the subasymptotic
elastic scattering data. At low energy it is compensated by the $f$-reggeon
contribution and at high energy by the rising term $P_1$ of  pomeron
contribution. Possibly this negative term effectively takes into
account the multipomeron exchanges contribution. We think that due
to complicated interference of the positive terms (from $F(x,Q^2)$
and $P_1(x,Q^2)$) with the negative term (from $P_2(x,Q^2)$), it is
possible to describe $\sigma_T^{\gamma^*p}(W,Q^2)$ and
$F_2^p(x,Q^2)$ without a $Q^2$-dependent  pomeron intercept.

In order to see how quickly the asymptotic regime is reached in
the DP model, we plot versus $W$ in Fig. 4 the ratios of
contributions in the cross-section $\sigma_T^{\gamma^*p}(W,Q^2)$ due
to the subasymptotic terms $F(W,Q^2)$ and $P_2(W,Q^2)$ to the
asymptotic one $P_1(W,Q^2)$ at some fixed values of $Q^2$. One
can see that the asymptotic domain (where $P_1$ dominates) occurs
for high $W$ and moreover, is shifted to even higher $W$'s while
$Q^2$ rises. This proves that it may be incorrect to
draw conclusions on $F_2^p$ in term of its asymptotic contribution.

\begin{center}
\epsfig{figure=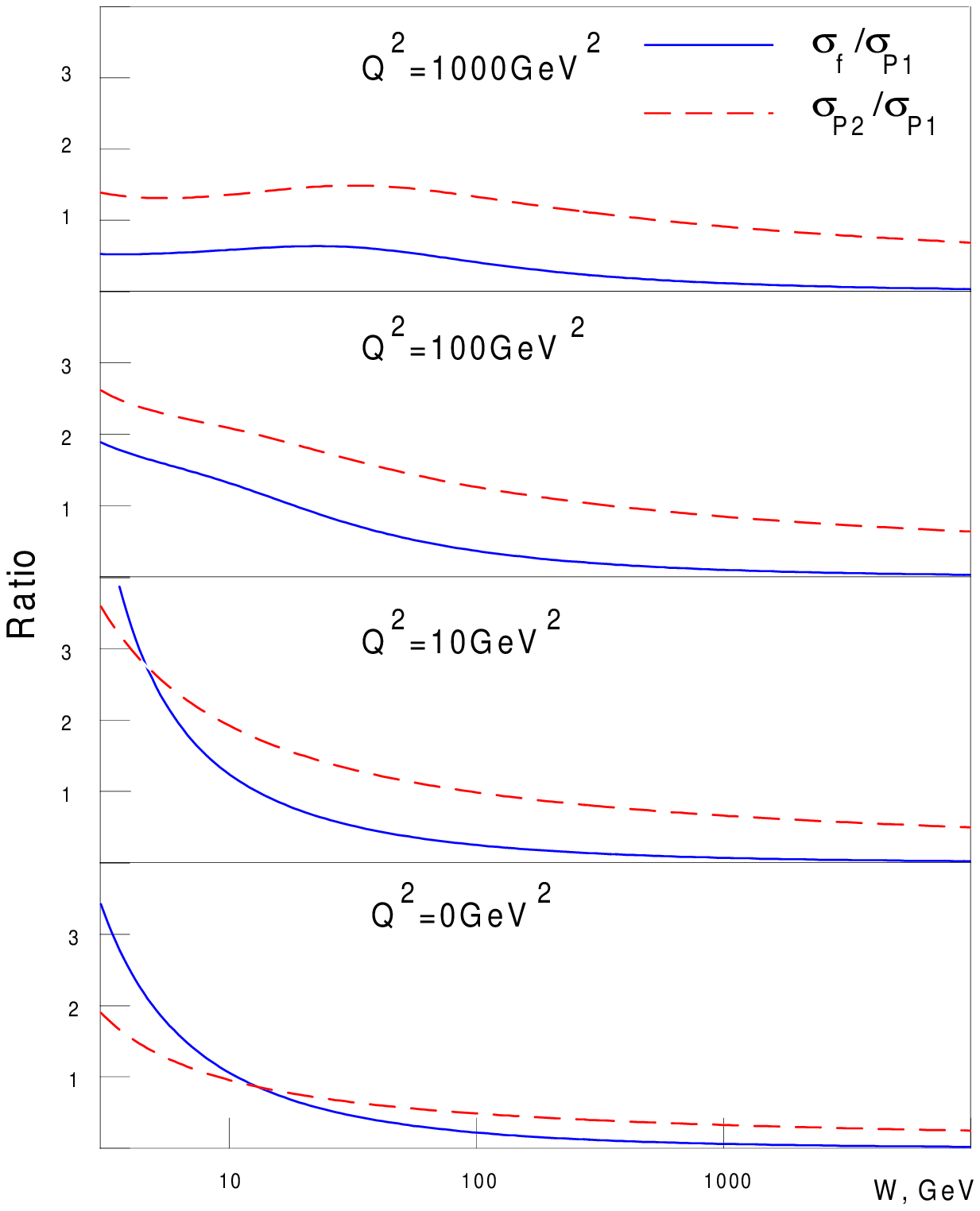,width=8cm}
\end{center}
\vskip .5cm
{\bf Figs.~4.} Ratios of $P_2$- and $F$- terms to $P_1$-term in
$\sigma_T^{\gamma^*p}(W,Q^2)$ versus $W$ at various fixed $Q^2$ as
indicated.
\vskip .5cm

{\bf 2.4 Comparison with other models}.

As noted above, models with $F_2^p\propto \ln(1/x)$ at $x\ll 1$,
can be interpreted as taking into account a Dipole Pomeron (with a
unit intercept) contribution to the SF. In Table 3 for two of them,
namely BH \cite{BH} and ScSp \cite{SS}, we compare the quality of
their description of the data (measured here with $\chi^2/{\rm (number\
of\ points)}$) with those obtained in the DP (present work) and
ALLM \cite{AL} models. The free parameters in all models were
(re)determined by fitting our set of data (for each model we
selected the kinematical region of $W,\,Q^2,\,x$ as indicated by
the authors). For ALLM and for our model we give the partial
$\chi^2$-s.
\vskip 0.3 cm

\noindent
{\bf Table 3.} Comparison of the quality of data description by
various models (measured here with $\chi^2$/(number of points)~).

\vskip 0.5cm
\begin{tabular}{|c|c|c|c|c|c|}\hline
Kinematical & Number of  & ALLM\cite{AL}   & ScSp\cite{SS}   &
BH\cite{BH}      & DP model     \\
region & points & & & & \\ \hline
  $W>3$ GeV & 1209    & 1.061     &  -    &   -  & 1.089     \\ \hline
  $0\leq Q^2\leq 350$ GeV$^2$ & 329    & 0.81    & 1.098     &  -    & 0.77
  \\
$W\geq 60$ GeV &   &      &      &    &  \\  \hline
$Q^2\geq 5$ GeV$^2$ & 227     & 0.89     &  -    & 1.04    & 0.79    \\
$x\leq 0.05$ & & & & &  \\ \hline
  \end{tabular}

\medskip
Because we describe $\sigma_T^{\gamma^*p}$ (or$F_2^p$) in the whole
kinematical region (the only restriction $W> 3$ GeV is imposed), we
can compare our model with the ALLM model \cite{AL} (where the only
restriction is $W>2$ GeV). We obtain a very small difference in the
whole region where data exist, though there is a quite different
trend outside the experimental range (see Figs.~2,3). Thus, future
experiments at lower $x$ should be crucial to test the existing
models and as a guide for constructing more sophisticated models.

\vskip 0.5cm

{\bf 2.5 ${\bf x-}$slope or} ${\bf {\partial ln
F_2^p(x,Q^2)/\partial ln(1/x)}}$.

The data suggest an interesting tendency in the behaviour of
$F_2^p(x,Q^2)$ at small $x$ and rising $Q^2$. It concerns the sharp
increase of $F_2$ as $x$ decreases for a large span of $Q^2$ values
(sometimes called the HERA effect). When $Q^2$ rises around
$Q^2\sim 200-500$ GeV$^2$, the fast growth of $F_2^p$ with
decreasing $x$ slows down and as $Q^2$ increases further is
reversed. This effect (let us call it damping of the HERA effect)
is very weak from the experimental point of view because lacking of
a sufficient number of data at high $Q^2$. Nevertheless, one can
see it (or simply constant that it does not contradict the
available experiments) {\it e.g.} in the Fig. 3d. In spite of the
very qualitative character of the experimental observation, the
following quantitative confirmation holds. The $x$-slope of the
proton SF
\begin{equation}\label{18}
B_x(x,Q^2)= \partial \ln{ ( F_2^p(x,Q^2) )}/\partial \ln (1/x),
\end{equation}
is strongly model dependent. As an example, we examine $B_x$ in
three models for which the behaviour of the rise of $F_2$ with
decreasing $x$ can accommodate all existing data~: the Dipole
Pomeron model (see above), the ALLM model \cite{AL} and the recent
model \cite{Lyon08} (hereafter labeled LKP). The asymptotic
behaviour (when $Q^2\to \infty$ and $x\ll1$) of this slope is
successively
$$ B_x^{(DP)} (x\ll 1,Q^2\to\infty)\approx \frac{1}{\ln(Q^2/x)},$$
$$ B_x^{(ALLM)}(x\ll 1,Q^2\to\infty)\approx \Delta(Q^2)=
a\left(1-1/\ln\ln(\frac{Q^2}{\Lambda ^2})\right),$$
$$ B_x^{(LKP)}(x\ll 1,Q^2\to\infty)\approx {1\over 2}
\sqrt{\gamma_1\ln\ln\frac{Q^2}{Q_0^2}\ / \ \ln\frac{x_0}{x}}\ ,$$
where $a, \Lambda, Q_0, \gamma _1, x_0$ are parameters of the
models. $B_x(x,Q^2)$ is plotted versus $Q^2$ for the DP
model in Fig. 5a for several $x$'s
as indicated. In Figs. 5b,c the $x-$slope for the ALLM and LKP models
is plotted for comparison. It should be noted that the $x-$slope,
at the largest experimental $Q^2$, is far from
asymptotics in all models. Our DP model predictions
differ strongly from those obtained in the ALLM model, where
the intercept $\Delta(Q^2)$ goes to a constant independent on $x$ at $Q^2\to
\infty$, and from those in the LKP model, where $B_x$ rises infinitely when
$Q^2\to \infty$. However in the domain of $Q^2$ slightly above the
existing data, the DP and LKP models are in qualitative agreement,
together predicting a decreasing $B_x$. They both predict the
damping of the HERA effect, illustrated by the presence of a
maximum when plotting $B_x$ versus $Q^2$ for a given low $x$
(Fig.5). New experimental data in this kinematical region would
certainly help to verify if this phenomenon does exist at $100
\lsim Q^2\lsim 1000GeV^2$.
\vskip 1.5cm
\begin{center}
\epsfig{figure=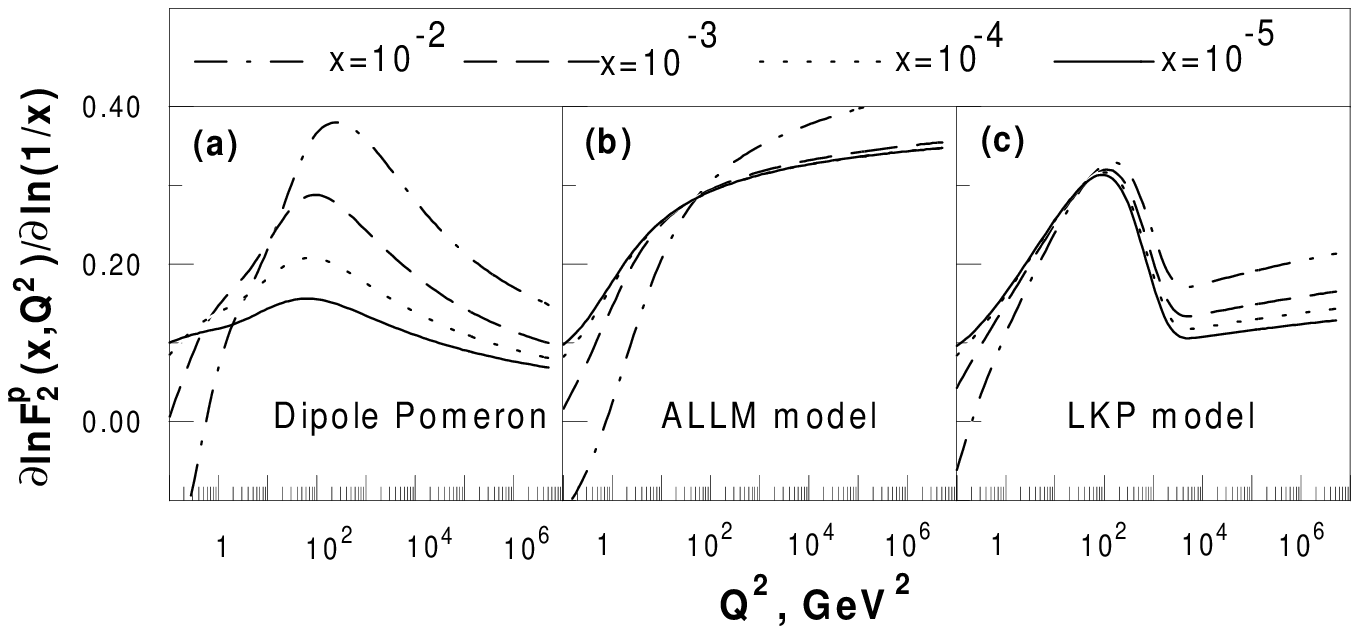,width=12cm}
\end{center}
\vskip .5cm
{\bf Figs.~5.} The $x$-slope of $F_2^p$ as function of $Q^2$ at the
four fixed $x$ in the DP, ALLM \cite{AL} and LKP \cite{Lyon08}
models.
\vskip .5cm

{\bf 2.6 "Effective intercept"} ${\bf \Delta^{(eff)}(x,Q^2)}.$

We would like to emphasize that an apparent contradiction of the
constant and "small" pomeron intercept (=1) in the DP model with
the "experimentally" established conclusion about the "high" (and
rising with $Q^2$) value of $\delta(Q^2)$, where $\delta$ is the
power in a small-$x$ behaviour of the SF, $F_2^p(x,Q^2)\propto
x^{-\delta(Q^2)}$, is not a real contradiction. On the one hand,
the conclusion is based on a simplified fit that takes into account
only the asymptotic contribution to the SF. On the second hand, the
"effective intercept" in the DP model rises with $Q^2$. To show
this we define this effective intercept $\Delta^{(eff)}$ by
rewriting $F_2^p$ in the general form
\begin{equation}
F_2^p(x,Q^2)=G(Q^2)\bigg (\frac{1}{x}\bigg)^{\Delta^{(eff)}(x,Q^2)}\  ,
\end{equation}
with
\begin{equation}
\Delta^{(eff)}(x,Q^2)=\alpha ^{(eff)}_{\cal P}(x,Q^2)-1\ .
\end{equation}

Note that $\Delta^{(eff)}(x,Q^2)=B_x(x,Q^2)$ only if
$\Delta^{(eff)}(x,Q^2)$ in Exp.(19) does not depend on $x$, in
general an effective intercept does not coincide with $x$-slope.

It is easy to obtain from (5-14) that in the DP model at
small fixed $x$ and $Q^2\to \infty$
$$F_2^p(x,Q^2)\approx {1\over 4\pi^2\alpha} Q^2G_1(Q^2)\ln W^2\approx
{1\over 4\pi^2\alpha} Q^2G_1(Q^2)\ln(\frac{Q^2}{x})$$
Identifying
with (19), one gets $G(Q^2)= {1\over 4\pi^2\alpha} Q^2G_1(Q^2)$ and
$$\Delta^{(eff)}(x,Q^2)\approx \frac{\ln\ln (Q^2/x)}{\ln(1/x)}\   . $$

Thus in the Dipole Pomeron model the effective intercept is
rising with rising $Q^2$ and is decreasing with decreasing $x$,
at least for large $Q^2$ and small $x$.
\bigskip
\vskip 0.5cm

{\bf 2.7 ${\bf Q-}$slope } or ${\bf \partial
F_2^p(x,Q^2)/\partial ln Q^2}$.

Recently, new low $x$ data from HERA have been reported
\cite{Cald,Bauer} and discussed \cite{AL,Lyon08,GoLeMa}, concerning
the logarithmic $Q^2$ derivative of $F_2^p$ (for brevity called
$Q-$slope)
\begin{equation}
B_Q(x,Q^2)=\frac{\partial F_2^p(x,Q^2)}{\partial \ln Q^2}.
\end{equation}
A "$Q-$slope effect" presented as a new phenomenon has been attributed
to these data, it is illustrated in Fig.6: the data
exhibit a peak at $Q^2_0\sim 1 - 5$ GeV$^2$. Also shown
in the figure, are the results of a calculation within our DP model,
which is a pure Regge one; quite a good agreement
with the data for both sides of the peak is obtained.
\vskip 1.5 cm
\begin{center}
\epsfig{figure=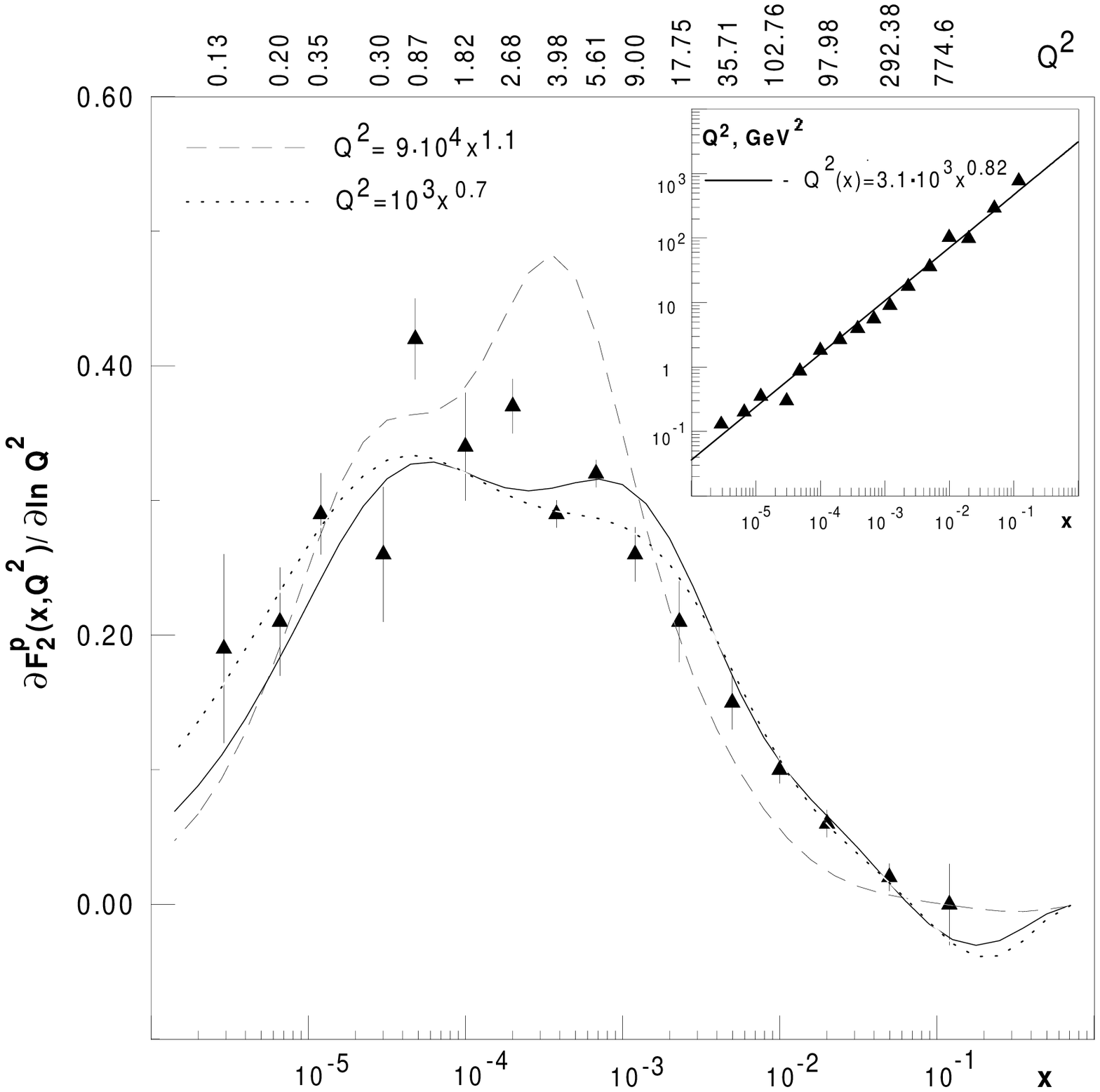,width=10cm}
\end{center}
\vskip .5cm
{\bf Figs.~6.} Slope $B_Q(x,Q^2)$. Experimental data from HERA and
results of calculation in the Dipole Pomeron model. Curves
correspond to the different choice of line $Q^2(x)$ (see the text).
\vskip 0.5 cm

The peak is currently interpreted as a transition region from a
Regge behaviour (at $Q^2\lsim Q^2_0$) to a perturbative QCD
regime (at $Q^2\gsim Q^2_0$). However we emphasize that such a
value of $Q_0^2$ is imposed by the specified selection of
experimental points, or in other words with \cite{Lyon08} by the
particular (experimentally constrained) path ($Q^2(x)$) chosen
on the surface representing $B_Q$ in the 3 dimensions space. Consequently,
an unbiased
determination of the transition region requires a study of the
submits of this surface and may yield $Q_0^2$, depending on $x$, and very
different from $1 - 5$ GeV$^2$ (it has been found $Q^2_0(x\ll 1)
\sim 40$ GeV$^2$ in \cite{Lyon08}).

In the mini-plot at the upper right corner of the Fig.~6, we show
the data positions in a ($x,Q^2$)-plane together with
a line $Q^2=3.1\cdot10^3\ x^{0.82}$ which is fitted to the data.
Solid line in the Fig.~6 corresponds to $B_Q$ calculated in the DP
model along this path.

Let us connect $x$ and $Q^2$ by some analytical dependence $Q^2=Q^2(x)$
that lies within a physical region on $(x,Q^2)$-plane. This region
is bounded by the condition
$$y=\frac{Q^2}{x(s-m_p^2)}\leq 1.$$
For HERA experiments, the c.m.s. energy is $\sqrt{s}\approx 300$
GeV and this condition writes $Q^2$ (in GeV$^2$) $< 9\cdot10^4\ x$.
As examples, we have calculated also $B_Q(x,Q^2)$ for two arbitrary
dependences satisfying the above condition~: $Q^2=9\cdot10^4\
x^{1.1}$ and $Q^2=10^3x\ ^{0.7}$. The results are given in Fig.~6
and show that the positions of the peaks in $x$ differ at least by
an order of magnitude. By an appropriate choice of the curve
$Q^2(x)$ the difference can be enforced. Thus, a peak indeed exists
but its position is strongly dependent on the choice of
experimental data. Undoubtedly, the $Q-$slope effect has to be
investigated in more details from the experimental and theoretical
points of view.

\vskip 1.cm
{\Large \bf Conclusion}
\vskip 0.5cm

In our opinion the most interesting and important message of that
paper is the following. All available data on the proton structure
function at $W>3$ GeV can be described in the framework of the
traditional Regge approach with a soft pomeron and an appropriate
$Q^2-$dependence of the residue function. It is not necessary for
this aim to use an high intercept similar to the "Born" hard BFKL
Pomeron intercept or a $Q^2$-dependent intercept. We find that the
main difficulty is the extension of the kinematical domain, where
the pomeron is successfully applied, from the small $x\ll 1$ to the
large $x\sim 1$ rather than the choice of the specified pomeron
singularity (in the $j$-plane) and its intercept. It means that the
subasymptotic contributions are extremely important not only at low
$W$ but also at HERA energy and even at more high energies (for the
DP model it is illustrated in the Figs.~4,5).

We note that in the DP model the "effective intercept" rises
infinitely when $Q^2$ rises and goes to zero when $x$ decreases.

At the same time, the model predicts a new phenomenon in the
behaviour of the slope $B_x(x,Q^2)$. Namely the observed rising growth of
the $x-$slope of $F_2^p$, at small $x$ and high $Q^2$, will come to
stop and then will begin to decrease at highest $Q^2$. This
phenomenon corresponding to a damping of the HERA effect
requires a further investigation.

\newpage
{\Large \bf Acknowledgments}
\vskip 0.5 cm
We are grateful to E.~Predazzi for reading the manuscript. We thank
M.~ Giffon and M.~Bertini for valuable discussions.


\begin{thebibliography}{99}
\bibitem{BGJPP} M. Bertini \etal, Rivista Nuovo Cimento {\bf 19}, 1 (1996).
\bibitem{Lev} A. Levy, {\it Low-$x$ physics at HERA}, DESY 97-013 (1997).
\bibitem{dglap} Yu.L. Dokshitzer, Sov. Phys. JETP {\bf 46}, 641 (1977);
L.V. Gribov, E.M. Levin, M.G. Ryskin, Phys. Rep. {\bf 100}, 1 (1983);
L.N. Lipatov in {\it Perturbative Quantum Chromodynamics}, edited by A.H.
Mueller (World Scientific, Singapore 1989); G. Altarelli, G. Parisi, Nucl.
Phys. B {\bf 126}, 298 (1977).
\bibitem{L1} E. Levin,
{\it Lessons and puzzles of DIS at low $x$ (high energy)}, TAUP 2435-97;
hep-ph/9706448 (1997).
\bibitem{L2} E. Levin,
{\it Where are the BFKL Pomeron and shadowing corrections in DIS~?},
DESY 97-171;
 hep-ph/9709226 (1997).
\bibitem{bfkl} Y.Y. Balitskii, L.N. Lipatov, Sov. Phys. JETP {\bf 28}, 822
(1978); E.A. Kuraev, L.N. Lipatov, V.S. Fadin, ibid {\bf 45}, 199 (1977);
L.N. Lipatov, ibid {\bf 63}, 904 (1986).
\bibitem{FL} V.S. Fadin, L.N. Lipatov,
{\it BFKL Pomeron in the next-to-leading approximation}, DESY 98-033;
hep-ph/9802290 (1998).
\bibitem{FDL} A.L. Ayala Filho, M.B. Gay Ducati, E.M. Levin,
{\it Scaling violation and shadowing corrections at HERA}, DESY
97-212; hep-ph/9710539 (1997).
\bibitem{GLM} E. Gotsman, E.M. Levin, U. Maor,
{\it A unitarity bound and the components of photon-photon interactions},
DESY 97-154; hep-ph/9708275 (1997).
\bibitem{GRV} M. Gl\"uk, E. Reya, A. Vogt, Zeit. Phys. C {\bf 67}, 433
(1995);
V. Barone, M. Genovese, N.N. Nicolaev, E. Predazzi, B.G. Zakharov,
Zeit. Phys. C {\bf 70}, 83 (1996).
\bibitem{MRS} A.D. Martin, R.G. Roberts, W.J. Stirling, Phys. Lett. B
{\bf 387}, 419 (1996).
\bibitem{CTEQ} H.L. Lai \etal (CTEQ Collaboration), Phys. Rev. D {\bf 51},
4763 (1996).
\bibitem{HLS} Z. Huang, H. Jung Lu, I. Sarcevic,
{\it Partonic picture of nuclear shadowing at small $x$}, AZPH-TH/97-07;
hep-ph/9705250 (1997).
\bibitem{Petr} V.A. Petrov, Nucl. Phys. (Proceedings Suppl.) A {\bf 54},
160 (1997).
\bibitem{CKMT} A. Capella \etal, Phys. Lett. B {\bf 349}, 561 (1995).
\bibitem{ALLM} H. Abramovicz \etal, Phys. Lett. B {\bf 269}, 465 (1991).
\bibitem{AL} H. Abramovicz, A. Levy,
{\it The ALLM parameterization of $\sigma_{tot}(\gamma^*,p)$ an
update}, DESY 97-251; hep-ph/9712415 (1997).
\bibitem{land94} P. V. Landshoff, {\it The two pomerons}, in {\it
Proceedings of the summer school on hadronic aspects of collider physics,
Zuoz, 1994}, edited by E.P. Locher (Villigen, PSI-Proceedings 94-01, 1994),
p. 135.
\bibitem{BGP} M. Bertini, M. Giffon, E. Predazzi, Phys. Lett. B {\bf 349},
561. (1995).
\bibitem{AY} K. Adel, F. Barreiro, F.J. Yndur\'{a}in, Nucl. Phys. B
{\bf 495}, 221 (1997).
\bibitem{M} E. Martynov, in {\it Proceedings of the workshop on soft
physics "Hadrons-94", Uzhgorod, Ukraine, September 1994}, edited by
G. Bugrij, L. Jenkovszky, E.Martynov (Kiev, 1994), p. 311; in {\it
Proceedings of the VI Blois Conference on Elastic and Diffractive
Scattering, Blois, France, June 1995}, edited by P. Chiapetta, M.
Haguenauer, J.Tr\^an Thanh V\^an (Editions Fronti\`eres 1996), p.
203.
\bibitem{JMP} L. Jenkovszky, E. Martynov, F. Paccanoni, {\it Regge
behavior of nucleon structure function.} PFPD 95/TH/21, Padova
University (1995).
\bibitem{BH} W. Buchm\"uller, D. Haidt,
{\it Double logarithmic scaling of the structure function $F_2$ at
small $x$},
 DESY 96-061; hep-ph/9605428 (1996).
\bibitem{SS} D. Schildknecht, H. Spiesberger,
{\it Generalized vector dominance and low $x$ inelastic
electron-proton scattering}, BI-TP97/25; hep-ph/9707447 (1997).
\bibitem{DGJLP} P. Desgrolard \etal, Phys. Lett. B {\bf 309}, 191 (1993).
\bibitem{DGLM} P. Desgrolard \etal, Nuovo Cimento A {\bf 107}, 637 (1994).
\bibitem{DGM} P. Desgrolard, M. Giffon, E. Martynov, Nuovo Cimento A
{\bf 110}, 537 (1997).
\bibitem{DL} A. Donnachie, P.V. Landshoff,  Zeit. Phys. C {\bf 61}, 139
 (1994).
\bibitem{HA} we thank H. Abramovicz for explaining comments relative to ref.
[17].
\bibitem{Cald} A. Caldwell, Invited talk in the DESY Theory
Workshop (October 1997).
\bibitem{Bauer} L. Bauerdick, {\it Proton structure function and
($\gamma^*,p$) cross-section at HERA}. Talk at the Max Planck workshop,
Heidelberg (1997).
\bibitem{Lyon08} P. Desgrolard, L. Jenkovszky, F. Paccanoni,
{\it Interpolating between soft and hard dynamics in deep inelastic
scattering}, LYCEN 9808; hep-ph/9803286 (1988); P. Desgrolard, L.
Jenkovszky, F. Paccanoni, {\it Small $x$, all $Q^2$ proton
structure function in DIS}, in {\it Proceedings of the 6$^{ th}$
workshop on Deep Inelastic Scattering and QCD}, Brussels, 1998 (in
press).
\bibitem{GoLeMa} E. Gotsman, E. Levin, U. Maor, {\it The F$_2$ slope and
shadowing corrections in DIS},
TAUP 2471/97; hep-ph/9712517 (1997).
\end{thebibliography}
\end{document}